\documentclass[12pt]{article}
\usepackage{epsf,amssymb}
\input{pdf1.tbl}
\input{pdf1.fig}
\newcommand{\DATE}  {\today \\ \strut}

\newcommand{\PPrtNo}
{
MSU-HEP-07101 \\
CERN-TH/2000-360
}
\newcommand{\TITLE}
{
Uncertainties of predictions from parton \\
distribution functions II: the Hessian method \strut
}
\newcommand{\AUTHORS}
{ J.\ Pumplin, D.\ Stump, R.\ Brock, D.\ Casey, J.\ Huston, \\
J.\ Kalk, H.L.\ Lai,$^a \strut$ W.K.\ Tung$^b \strut$
}
\newcommand{\INST}
{ Department of Physics and Astronomy \\
         Michigan State University \\
         East Lansing, MI 48824 \\

\vspace{2ex}

         $^a$ Ming-Hsin Institute of Technology \\
         Hsin-Chu, Taiwan

\vspace{2ex}

         $^b$ Theory Division, CERN \\
         Geneva, Switzerland
}
\newcommand{\ABSTRACT}
{ We develop a general method to quantify the uncertainties of parton
distribution functions and their physical predictions, with emphasis on
incorporating all relevant experimental constraints.  The method uses
the Hessian formalism to study an effective chi-squared function that
quantifies the fit between theory and experiment.  Key ingredients are
a recently developed iterative procedure to calculate the Hessian matrix
in the difficult global analysis environment, and the use of parameters
defined as components along appropriately normalized eigenvectors.  The
result is a set of $2d$ Eigenvector Basis parton distributions (where
$d \! \approx \! 16$ is the number of parton parameters) from which
the uncertainty on any physical quantity due to the uncertainty in
parton distributions can be  calculated.  We illustrate the method by
applying it to calculate uncertainties of gluon and quark distribution
functions, $W$ boson rapidity distributions, and the correlation between
$W$ and $Z$ production cross sections. }

\newpage
\input{pdf1.pre}
\begin{document}

\begin{titlepage}

\begin{tabular}{l}
\noindent\DATE
\end{tabular}
\hfill
\begin{tabular}{l}
\PPrtNo
\end{tabular}

\vspace{1cm}

\begin{center}
\renewcommand{\thefootnote}{\fnsymbol{footnote}}
{
\LARGE \TITLE
}

\vspace{1.25cm}
{\large  \AUTHORS}

\vspace{1.25cm}

\INST
\end{center}

\vfill

\ABSTRACT                 

\vfill

\newpage
\end{titlepage}

\renewcommand{\thefootnote}{\alph{footnote}}   
\setcounter{footnote}{0}


\tableofcontents
\newpage
%



\section{Introduction}
\label{sec:Intro}

The partonic structure of hadrons plays a fundamental role in
elementary particle physics.  Interpreting experimental data according to
the Standard Model (SM), precision measurement of SM parameters, and searches
for signals of physics beyond the SM, all rely on the parton picture of
hadronic beam particles that follows from the factorization theorem of Quantum
Chromodynamics (QCD). The parton distribution functions (PDFs) are
nonperturbative---and hence at present uncalculable---functions of momentum
fraction $x$ at a low momentum transfer scale $Q_0$. They are determined
phenomenologically by a global analysis of experimental data from a wide
range of hard-scattering processes, using perturbative QCD to
calculate the hard scattering and to determine the dependence of the PDFs on
$Q$ by the renormalization-group based evolution equations.

Considerable progress has been made in several parallel efforts to improve our
knowledge of the PDFs \cite{cteq5,MRS,GRV}, but many problems remain open.  In
the conventional approach, specific PDF sets are constructed to represent the
``best estimate'' under various input assumptions, including selective
variations of some of the parameters \cite{Cteq4,MRSs,CteqGlu}.  From these
results, however, it is impossible to reliably assess the uncertainties of the
PDFs or, more importantly, of the physics predictions based on them.  The need
to quantify the uncertainties for precision SM studies and New Physics
searches in the next generation of collider experiments has stimulated much
interest in developing new approaches to this problem \cite{TeVIIwks,LHCwks}.
Several attempts to quantify the uncertainties of PDFs in a systematic
manner have been made recently \cite{Alehkin,Botje,Barone,GKK,UncRunII}.

The task is difficult because of the diverse sources of experimental and
theoretical uncertainty in the global QCD analysis.
In principle, the natural framework for studying uncertainties is that of the
likelihood function \cite{GKK,Callan,Ball}. If all experimental measurements
were available in the form of mutually compatible probability functions for
candidate theory models, then the combined likelihood function would
provide the probability distribution for the possible PDFs
that enter the theory.  From this, all physical predictions and their
uncertainties would follow. Unfortunately, such ideal likelihood functions
are rarely available from real experiments.  To begin with, most published
data sets used in global analysis provide only effective errors
in uncorrelated form, along with a single overall normalization
uncertainty.
Secondly, published errors for some well-established experiments
appear to fail standard statistical tests, {\textit e.g.}, the
$\chi^{2}$ per degree of freedom may deviate significantly from $1.0$,
making the data set quite ``improbable''.  In addition,
when the few experiments that are individually amenable to likelihood
analysis are examined together, they appear to demand mutually
incompatible PDF parameters.  A related problem is that
the theoretical errors are surely highly correlated and by definition poorly
known.  All these facts of life make the idealistic approach
impractical for a real-world global QCD analysis.

The problems that arise in combining a large number of diverse experimental
and theoretical inputs with uncertain or inconsistent errors are similar to
the problems routinely faced in analyzing systematic errors within
experiments, and in averaging data from measurements that are
marginally compatible \cite{PDG}.  Imperfections of data sets in the form
of unknown systematic errors or unusual fluctuations---or both---are
a common occurrence.  They need not necessarily destroy the value of those
data sets in a global analysis; but we must adapt and expand the
statistical tools we use to analyze the data, guided by reasonable physics
judgement.

In this paper we develop a systematic procedure to study the uncertainties of
PDFs and their physics predictions, while incorporating all the experimental
constraints used in the previous CTEQ analysis \cite{cteq5}.
An effective $\chi^{2}$
function, called $\chi_{\mathrm{global}}^2$, is used not only to extract the
``best fit'', but also to explore the neighborhood of the global minimum in
order to quantify the uncertainties, as is done in the classic Error Matrix
approach.  Two key ingredients make this possible:
(i) a recently established iterative procedure \cite{Paper0} that
yields a reliable calculation of the Hessian matrix in the complex global
analysis environment; and
(ii) the use of appropriately normalized eigenvectors
to greatly improve the accuracy and utility of the analysis.

The Hessian approach is based on a quadratic approximation to
$\chi_{\mathrm{global}}^2$ in the neighborhood of the minimum that defines the
best fit.  It yields a set of PDFs associated with the eigenvectors
of the Hessian, which characterize the PDF parameter space in the neighborhood
of the global minimum in a \emph{process-independent} way.  In a companion
paper, referred to here as LMM \cite{LMM}, we present a
complementary process-dependent method that studies
$\chi_{\mathrm{global}}^2$ as a function of whatever specific physical variable
is of interest.  That approach is based on the Lagrange Multiplier
method \cite{Paper0}, which does not require a quadratic approximation to
$\chi_{\mathrm{global}}^2$,
and hence is more robust; but, being focused on a single variable (or a
few variables in a generalized formulation), it does not provide complete
information about the neighborhood of the minimum.  We use the LM method
to verify the reliability of the Hessian calculations, as discussed in
Sec.~\ref{sec:PhysApp}.  Further tests of the
quadratic approximation are described in Appendix \ref{sec:QuadApprox}.

The outline of the paper is as follows.
In Sec.~\ref{sec:Global} we summarize the global analysis that underlies
the study, and define the function $\chi_{\mathrm{global}}^2$ that plays the
leading role.
In Sec.~\ref{sec:Hessian} we in explore the quality of fit in the neighborhood
of the minimum.ÿWe derive the Eigenvector Basis sets, and show how they
can be used to calculate the uncertainty on any quantity that depends on the
parton distributions.
In Sec.~\ref{sec:UncPDF} we apply the formalism to derive uncertainties
of the PDF parameters and of the PDFs themselves.
In Sec.~\ref{sec:PhysApp} we illustrate the method further by finding the
uncertainties on predictions for the rapidity distribution of $W$
production, and the correlation between $W$ and $Z$ production cross sections.
We summarize our results in Sec.~\ref{sec:Conclude}.
Two appendices provide details on the estimate of overall tolerance for the
effective  $\chi_{\mathrm{global}}^2$ function, and on the validity of the
quadratic approximation inherent in the Hessian method.
Two further appendices supply explicit tables of the coefficients that define
the best fit and the Eigenvector Basis sets.
The mathematical methods used here have been described in detail
in \cite{Paper0}.  Some preliminary results have also appeared
in \cite{TeVIIwks,LHCwks}.




\section{Global QCD Analysis and Effective Chi-squared}
\label{sec:Global}

Global $\chi^2$ analysis is a practical and effective way to combine a
large number of experimental constraints.  In this section, we describe
the main features of the global QCD analysis, and explain how we quantify
its uncertainties through the behavior of $\chi_{\mathrm{global}}^{2}$.

\subsection{Experimental and theoretical inputs}
\label{sec:Inputs}

We use the same experimental input as the CTEQ5
analysis \cite{cteq5}: 15 data sets on neutral-current and
charged-current deep inelastic scattering (DIS), Drell-Yan lepton pair
production (DY), forward/backward lepton asymmetry from $W$ production,
and high $p_{T}$ inclusive jets, as listed in Table \ref{tbl:DatSet} of
Appendix \ref{sec:Tolerance}.  The total number of data points is $1295$
after cuts, such as $Q > 2  \, {\rm GeV}$ and $W > 4 \, {\rm GeV}$ in DIS,
designed to reduce the influence of power-law suppressed corrections
and other sources of theoretical error.
The experimental precision and the information available on systematic
errors vary widely among the experiments, which presents difficulties
for the effort to quantify the uncertainties of the results.

The theory input is next-to-leading-order (NLO) perturbative QCD.
The theory has systematic uncertainties due to uncalculated higher-order
QCD corrections, including possible resummation effects near kinematic
boundaries; power-suppressed corrections; and nuclear effects
in neutrino data on heavy targets.  These uncertainties---even more than
the experimental ones---are difficult to quantify.

The theory contains free parameters $\{a_i\}=\{a_1,\ldots,a_d\}$ defined
below that characterize the nonperturbative input to the analysis.
Fitting theory to experiment determines these $\{a_i\}$ and thereby the PDFs.
The uncertainty of the result due to experimental and theoretical errors
is assessed in our analysis by an assumption on the permissible range of
$\Delta \chi^2$ for the fit, which is discussed in Sec.~\ref{sec:Neighborhood}.

\subsection{Parametrization of PDFs}
\label{sec:Parametrization}

The PDFs are specified in a parametrized form at a fixed low energy scale
$Q_{0}$, which we choose to be $1 \, {\rm GeV}$.  The PDFs at all higher $Q$
are determined from these by the NLO perturbative QCD evolution equations.
The functional forms we use are
\begin{equation}
f(x,Q_{0}) = A_{0} \, x^{A_{1}} \, (1-x)^{A_{2}} \,
(1+A_{3} \, x^{A_{4}})
\label{eq:param}
\end{equation}
with independent parameters for parton flavor combinations
$u_{v} \equiv u - \bar{u}$, $d_{v} \equiv d - \bar{d}$, $g$, and
$\bar{u}+\bar{d}\,$. We assume
$s = \bar{s} = 0.2 \, (\bar{u} + \bar{d})$ at $Q_0$.
A somewhat different parametrization for the $\bar{d}/\bar{u}$ ratio is
adopted to better fit the current data:
\begin{equation}
\bar{d}(x,Q_{0})/\bar{u}(x,Q_{0}) =
B_{0}\,x^{B_{1}}\,(1-x)^{B_{2}} \; + \;(1+B_{3}x)\,(1-x)^{B_{4}} \;.
\label{eq:dou}
\end{equation}
The specific functional forms are not crucial, as long as the parametrization
is flexible enough to include the behavior of the
true---but unknown---PDFs at the level of accuracy to which
they can be determined.
The parametrization should also provide additional flexibility to model
the degrees of freedom that remain indeterminate.
On the other hand, too much flexibility in the
parametrization leaves some parameters poorly determined
at the minimum of $\chi^{2}$.  To avoid that problem, some parameters
in the present study were frozen at particular values.

The number of free parameters has increased over the years, as the accuracy
and diversity of the global data set has gradually improved.  A useful feature
of the Hessian approach is the feedback it provides to aid in refining the
parametrization, as we discuss in Sec.\ \ref{sec:UncParamAi}. The current
analysis uses a total of $d \! = \! 16$ independent parameters, referred to
generically as $\{a_i\}$.  Their best-fit values, together with the fixed
ones, are listed in Table~\ref{tbl:AiMatrix} of Appendix \ref{sec:S0Table}.
(Some of the fit parameters $a_i$ are defined by simple
functions of their related PDF shape parameters $A_i$ or $B_i$, as indicated
in the table, to keep their relevant magnitudes in a similar range, or to
enforce positivity of the input PDFs, {\textit etc}.)  The set of fit parameters
$\{a_i\}$ could also include parameters associated with correlated experimental
errors, such as an unknown normalization error that is common to all of
the data points in a particular experiment; however, such parameters were kept
fixed for simplicity in this initial study.  The QCD coupling was similarly
fixed at $\alpha_s(M_Z) = 0.118\,$.

\subsection{Effective chi-squared function}
\label{sec:ChiSqFn}

Our analysis is based on an effective global chi-squared function
that measures the quality of the fit between theory and experiment:
\begin{equation}
\chi_{\mathrm{global}}^{2} = \sum_{n} \, w_{n} \, \chi_{n}^{2}
\label{eq:Chi2global}
\end{equation}
where $n$ labels the $15$ different data sets.

The weight factors $w_{n}$ in (\ref{eq:Chi2global}), with default value $1$,
are a generalization of the selection process that must begin any global
analysis, where one decides which data sets to include ($w\!=\!1$) or exclude
($w\!=\!0$). For instance, we include neutrino DIS data (because it contains
crucial constraints on the PDFs, although it requires model-dependent nuclear
target corrections); but we exclude direct photon data (which would help to
constrain the gluon distribution, but suffers from delicate sensitivity to
$k_\perp$ effects from multiple soft gluon emission). The $w_n$ can be used
to emphasize particular experiments that provide unique physical information,
or to de-emphasize experiments when there are reasons to suspect unquantified
theoretical or experimental systematic errors ({\textit e.g.} in comparison to
similar experiments).  Subjectivity such as this choice of weights is not
covered by Gaussian statistics, but is a part of the more general
Bayesian approach; and is in spirit a familiar aspect of estimating systematic
errors within an experiment, or in averaging experimental results that are
marginally consistent.

The generic form for the individual contributions in
Eq.~(\ref{eq:Chi2global}) is
\begin{equation}
\chi_n^{2} = \sum_{I}
\left(\frac{D_{nI} - T_{nI}}{\sigma_{nI}}\right)^{2}
\label{eq:Ch2generic}
\end{equation}
where $T_{nI}$, $D_{nI}$, and $\sigma_{nI}$ are the theory value, data value,
and uncertainty for data point $I$ of data set (or ``experiment'') $n$.
In practice, Eq.~(\ref{eq:Ch2generic})
is generalized to include correlated errors such as overall normalization
factors; or even the full experimental error correlation matrix if it is
available \cite{LMM}.

The value of $\chi_{\mathrm{global}}^{2}$ depends on the PDF set, which we
denote by $S$.
We stress that $\chi_{\mathrm{global}}^{2}(S)$ is an ``effective $\chi^2\,$'',
whose purpose is to measure how well the data are fit by the theory when
the PDFs are defined by the parameter set $\{a_i(S)\}$.  We use
$\chi_{\mathrm{global}}^{2}(S)$ to study how
the quality of fit varies with the PDF parameters; but we do not assign
\textit{a priori} statistical significance to specific values of
it---\textit{e.g.}, in the manner that would be appropriate to an ideal
chi-squared distribution---since
the experimental and theoretical inputs are often far from being ideal,
as discussed earlier.

\subsection{Global minimum and its neighborhood}
\label{sec:Neighborhood}

Having specified the effective $\chi^2$ function, we find the parameter
set that minimizes it to obtain a ``best estimate'' of the true PDFs.
This PDF set is denoted by $S_0$.\footnote{It is very similar to the
CTEQ5M1 set \cite{cteq5}, with minor differences arising from the improved
parametrization (\ref{eq:dou}) for $\bar{d}/\bar{u}$.}  The parameter
values that characterize $S_0$ are listed in Table~\ref{tbl:AiMatrix} of
Appendix \ref{sec:S0Table}.

To study the uncertainties, we must explore the variation  of
$\chi_{\mathrm{global}}^{2}$ in the neighborhood of its minimum, rather
than focusing only on $S_0$ as has been done in the past.
Moving the parameters away from the minimum increases
$\chi_{\mathrm{global}}^{2}$ by an amount $\Delta \chi^2_{\mathrm{global}}$.
It is natural to define the relevant neighborhood of the global minimum as
\begin{equation}
\Delta \chi_{\mathrm{global}}^2 \, \le \, T^2
\label{eq:tolerance}
\end{equation}
where $T$ is a \emph{tolerance parameter}. The Hessian formalism developed in
the next section (Sec.~\ref{sec:Hessian}) provides a reliable and efficient
method of calculating the variation of all predictions of PDFs in this
neighborhood, as long as $T$ is within the range where a quadratic expansion
of $\chi_{\mathrm{global}}^{2}$ in terms of the PDF parameters is adequate.

In order to quantify the uncertainties of physical predictions that depend on
PDFs, one must choose the tolerance parameter $T$ to correspond to the region
of ``acceptable fits''. Broadly speaking, the order of magnitude of $T$ for
our choice of $\chi_{\mathrm{global}}^{2}$ is already suggested by
self-consistency considerations:  Our fundamental assumption---that the 15
data sets used in the global analysis are individually \emph{acceptable} and
mutually \emph{compatible}, in spite of departures from ideal statistical
expectations exhibited by some of the individual data sets, as well as signs
of incompatibility between some of them if the errors are interpreted
according to strict statistical rules \cite{GKK}---must, in this effective
$\chi^2$ approach, imply a value of $T$ substantially larger than that of
ideal expectations. More quantitatively, estimates of $T$ have been carried
out in the companion paper LMM \cite{LMM}, based on the comparison of
$\Delta\chi_{\mathrm{global}}^{2}$ with detailed studies of experimental
constraints on specific physical quantities. The estimates of $T$ will be
discussed more extensively in Sec.~\ref{sec:PhysApp}, where applications are
presented, and in Appendix \ref{sec:Tolerance}.  For the development of the
formalism in the next section, it suffices to know that {\it (i)} the order of
magnitude of these estimates is
\begin{equation}
T \approx 10 \; {\rm to} \; 15 \; ;
\label{eq:Tapprox100}
\end{equation}
and {\it (ii)} the master formulas given in Sec.~\ref{sec:MasterEq} show that all
uncertainties are proportional to $T$, so that results derived for a particular
value of $T$ can easily be scaled to other estimates of $T$ if desired.




\section{The Hessian formalism}
\label{sec:Hessian}

The most efficient approach to studying uncertainties in a global analysis
of data is through a quadratic expansion of the $\chi^2$ function about its
global minimum.\footnote{The Lagrange multiplier method \cite{Paper0,LMM}
is a complementary approach that avoids the quadratic approximation.}
This is the well known Error Matrix or Hessian method.  Although the method
is standard, its application to PDF analysis has, so far, been hindered by
technical problems created by the complexity of the theoretical and
experimental inputs.  Those technical problems have recently been
overcome \cite{Paper0}.

\figIllustrate  

The Hessian matrix is the matrix of second derivatives of $\chi^2$ at the
minimum.
In our implementation, the eigenvectors of the Hessian matrix
play a central role.  They are used both for accurate evaluation of the
Hessian itself, via the iterative method of \cite{Paper0}, and to produce
an Eigenvector Basis set of PDFs from which uncertainties of all physical
observables can be calculated.  These basis PDFs provide an optimized
representation of the parameter space in the neighborhood of the
minimum $\chi^2$.

The general idea of our approach is illustrated conceptually in
Fig.~\ref{fig:Illustrate}.  Every PDF set $S$ corresponds to a point in the
$d$-dimensional PDF parameter space.  It can be specified by the original
parton shape
parameters $\{a_i(S)\}$ defined in Sec.~\ref{sec:Parametrization}, as
illustrated in Fig.~\ref{fig:Illustrate}(a); or by the Eigenvector Basis
coordinates $\{z_k(S)\}$, which specify the components of $S$ along the
Eigenvector Basis PDFs that will be introduced in
Sec.~\ref{sec:Eigenvectors}, as illustrated in Fig.~\ref{fig:Illustrate}(b).
The solid points in both (a) and (b) represent the basis PDF sets.

\subsection{Quadratic approximation and the Hessian matrix}
\label{sec:HessianMtx}

The standard error matrix approach begins with a Taylor series expansion of
$\chi_{\mathrm{global}}^{2}(S)$ around its minimum $S_0$, keeping
only the leading terms. This produces a quadratic form
in the displacements from the minimum:
\begin{equation}
\Delta \chi^{2} \, = \, \chi^{2}-\chi_{0}^{\,2} \, = \,
\sum_{i=1}^d \sum_{j=1}^d \,
H_{ij} \, (a_{i} - a^{0}_{i}) \, (a_{j} - a^{0}_j)
\label{eq:taylor}
\end{equation}
where $\chi_{0}^{\, 2} = \chi^2(S_0)$
is the value at the minimum, $\{a_j^0\} = \{a_j(S_0)\}$ is its location, and
$\{a_j\} = \{a_j(S)\}$.
We have dropped the subscript ``global'' on $\chi^2$ for
simplicity.  We also suppress the PDF argument $(S)$ in $\chi^2$ and $\{a_i\}$
here and elsewhere for brevity.

The Hessian matrix $H_{ij}$ has a complete set of orthonormal
eigenvectors $v_{ik}$ defined by
\begin{eqnarray}
\sum_{j=1}^{d} \, H_{ij}\,v_{jk} &=&\epsilon_{k}\,v_{ik}
\label{eq:Hevec}
\\
\sum_{i=1}^d \, v_{i\ell}\,v_{ik} &=&\delta_{\ell k} \;,
\label{eq:orthonormal}
\end{eqnarray}
where $\{\epsilon_{k}\}$ are the eigenvalues and $\delta_{\ell k}$ is the
unit matrix.  
Displacements from the minimum are conveniently expressed
in terms of the eigenvectors by
\begin{equation}
a_{i} - a^{0}_i = \sum_{k=1}^d \, v_{ik} \, s_k \, z_{k} \; ,
\label{eq:vij}
\end{equation}
where scale factors $s_k$
are introduced to normalize the new parameters $z_k$ such that
\begin{equation}
\Delta \chi^{2} \, = \,\sum_{k=1}^d \, z_{k}^{\,2} \; .
\label{eq:vij2}
\end{equation}
With this normalization,
the relevant neighborhood (\ref{eq:tolerance}) of the global
minimum corresponds to the interior of a hypersphere of radius $T$:
\begin{equation}
\sum_{k=1}^d \, z_k^{\,2} \, \le \, T^2 \; .
\label{eq:sphere}
\end{equation}
In the ideal quadratic approximation, the scale factors $s_k$ would be equal to
$\sqrt{1/\epsilon_k}\,$.  However, if $\chi^{2}$ is not perfectly quadratic,
then $s_k$ differs somewhat from $\sqrt{1/\epsilon_k}\,$, as explained in
Appendix \ref{sec:QuadApprox}.

The transformation (\ref{eq:vij}) is illustrated conceptually in
Fig.~\ref{fig:Illustrate}, where $\{k,\ell\}$ label two of the eigenvector
directions. One of the eigenvectors $\mathbf{v}_\ell$ is shown both in the
original parameter basis and in the normalized eigenvector basis.

\subsection{Eigenvalues of the Hessian matrix}
\label{sec:EigenValues}

The square of the distance in parameter space from the minimum of
$\chi^{2}$ is
\begin{eqnarray}
\sum_{i=1}^d \, (a_i - a_i^0)^{2} \, = \,
\sum_{k=1}^d \, (s_k \, z_k)^{2}
\label{eq:lengthsq}
\end{eqnarray}
by (\ref{eq:orthonormal})--(\ref{eq:vij}).
Because $s_k \approx \sqrt{1/\epsilon_k}$,
an eigenvector with large eigenvalue $\epsilon_k$ therefore corresponds to
a ``steep direction'' in $\{a_i\}$ space, {\textit i.e.}, a direction in which
$\chi^2$ rises rapidly, making the parameters tightly constrained by the
data. The opposite is an eigenvector with small $\epsilon_k$, which
corresponds to a ``shallow direction'', for which the criterion
$\Delta \chi^2 \le T^2$ permits considerable motion---as is the
case for $\mathbf{v}_\ell$ illustrated in Fig.~\ref{fig:Illustrate}.

The distribution of eigenvalues, ordered from largest to smallest,
is shown in Fig.~\ref{fig:Eigenvalue}.
Interestingly, the distribution
is approximately linear in $\, \log \epsilon \,$.
The eigenvalues span an enormous range, which is understandable because
the large global data set includes powerful constraints---particularly on
combinations of parameters that control the quark distributions at moderate
$x$---leading to steep directions; while free parameters have purposely
been added to (\ref{eq:param})--(\ref{eq:dou}) to the point where
some of them are at the edge of being unconstrained by the data, leading to
shallow directions.

\figEigenvalue  

Fig.~\ref{fig:Eigenvalue} also shows how the range of eigenvalues expands or
contracts if the number of adjustable parameters is changed:  the 16-parameter
fit is the standard one used in most of this paper; the 18-parameter fit
is defined by allowing $A_1^{u_v} \ne A_1^{d_v}$ and $A_3^{g} \ne 0$ with
$A_4^{g} = 6$; the 13-parameter fit is defined by
$A_4^{u_v} = A_4^{d_v} = 1$ and $A_3^{\bar{d} + \bar{u}} = 0$.
The range spanned by the eigenvalues increases with the
dimension $d$ of the parameter space
(roughly as $\, \log(\epsilon_1/\epsilon_d) \propto d^{0.4}\,$).

The large ($10^{6} \! : \! 1$) range spanned by the eigenvalues makes the
smaller eigenvalues and their eigenvectors very sensitive to fine details
of the Hessian matrix, making it difficult to compute $H_{ij}$ with
sufficient accuracy.  This technical problem hindered the use of Hessian
or Error Matrix methods in global QCD analysis in the past.  The problem
has been tamed by an iterative method introduced in  \cite{Paper0}, which
computes the eigenvalues and eigenvectors by successive approximations that
converge even in the presence of numerical noise and non-quadratic
contributions to $\chi^2\,$.\footnote{The
iterative algorithm is implemented as an extension to the widely used CERNLIB
program {\footnotesize MINUIT} \cite{minuit}.  The code is available at
http://www.pa.msu.edu/${\scriptstyle \sim}$pumplin/iterate/.}

The Hessian method relies on the quadratic approximation (\ref{eq:taylor}) to
the effective $\chi^2$ function.  We have extensively
tested this approximation in the neighborhood of interest
by comparing it with the exact $\chi^2_\mathrm{global}$.
The results are satisfactory, as shown in
Appendix \ref{sec:QuadApprox}, which also explains how the approximation
is improved by adjusting the scale factors $s_k$ for the shallow directions.

\subsection{PDF Eigenvector Basis sets $S_{\ell}^{\pm }$}

\label{sec:Eigenvectors}

The $k^{\rm th}$ eigenvector of the Hessian matrix has component $v_{ik}$ along
the $i^{\rm th}$ direction in the original parameter space, according to
(\ref{eq:Hevec}).  Thus $v_{ik}$ is the orthogonal matrix that transforms
between the original parameter basis and the eigenvector basis.  For our
application, it is more convenient to work with coordinates $\{z_{k}\}$
that are normalized by the scale factors $\{s_k\}$ of
(\ref{eq:vij}),
rather than the ``raw'' coordinates of the eigenvector basis.
Thus we use the matrix
\begin{equation}
M_{ik} \equiv v_{ik} \, s_{k}
\label{eq:Mik}
\end{equation}
rather than $v_{ik}$ itself.
$M_{ik}$ defines the transformation between the two descriptions that are
depicted conceptually in Fig.~\ref{fig:Illustrate}:
\begin{equation}
a_{i} - a^{0}_i = \sum_{k=1}^d \, M_{ik} \, z_{k} \; .
\label{eq:MikTransform}
\end{equation}
It contains information about the physics in the global fit, together with
information related to the choice of parametrization, and is a
good object to study for insight into how the parametrization might be
improved, as we discuss in Sec.~\ref{sec:UncParamAi}.

The eigenvectors provide an optimized orthonormal basis in the PDF parameter
space, which leads to a simple parametrization of the parton distributions
in the neighborhood of the global minimum $S_{0}$.  In the remainder of
this section, we show how to construct these \emph{Eigenvector Basis} PDFs
$\{S_{\ell }^{\pm },\ell =1,\ldots,d\}\,$; and in the following subsection,
we show how they can be used to calculate the uncertainty of any desired
variable $X(S)$.

The \emph{Eigenvector Basis} sets $S_{\ell }^{\pm }$ are defined by
displacements of a standard magnitude $t$ ``up'' or ``down'' along each of
the $d$ eigenvector directions.  Their coordinates in the $z$-basis are thus
\begin{equation}
z_{k}(S_{\ell }^{\pm })=\pm \,t\,\delta_{k\ell } \; .
\label{eq:eigenvectork-z}
\end{equation}
More explicitly, $S_{1}^{+}$ is defined by
$\left(z_1,\dots,z_d\right) = \left(t, 0,\dots,0\right)$, \textit{etc}.
We make displacements in both directions along each eigenvector to
improve accuracy; which direction is called ``up'' is totally arbitrary.
As a practical matter, we choose $t \! = \! 5$ for the displacement distance.
This choice improves the accuracy of the quadratic approximation by working
with displacements that have about the same size as those needed in
applications.\footnote{The value chosen for $t$ is somewhat smaller than
the typical $T$ given in (\ref{eq:Tapprox100}) because in applications, the
component of displacement along a given eigenvector direction is generally
considerably smaller than the total displacement.}

The $\{a_i\}$ parameters that specify the Eigenvector Basis sets
$S_{\ell }^{\pm }$ are given by
\begin{equation}
a_{i}(S_{\ell }^{\pm }) \, - \, a_{i}^{0} \, = \,
\pm \, t \, M_{i\ell }
\label{eq:eigenvectork-a}
\end{equation}
by (\ref{eq:MikTransform})--(\ref{eq:eigenvectork-z}).  Hence
\begin{equation}
a_{i}(S_{\ell}^{+}) - a_{i}(S_{\ell}^{-}) \, = \, 2 \, t \, M_{i\ell} \; .
\label{eq:DiffEq}
\end{equation}
Interpreted as a difference equation, this shows directly that the element
$M_{i\ell}$ of the transformation matrix is equal to the gradient of
parameter $a_{i}$ along the direction of $z_{\ell }$.\footnote{Technically,
we calculate the orthogonal matrix $v_{ij}$ using displacements that give
$\Delta \chi ^{2}\simeq 5$ where the iterative procedure \cite{Paper0}
converges well.  The eigenvectors are then scaled up by an amount that is
adjusted to make $\Delta \chi ^{2}=25$ exactly for each $S_{k}^{\pm }$ to
improve the quadratic approximation.
\label{fn:scale}}

Basis PDF sets along two of the eigenvector directions are illustrated
conceptually in Fig.~\ref{fig:Illustrate}~(a) and (b) as solid points
displaced from the global minimum set $S_{0}$.
The coefficients that specify
all of the sets $S_{\ell }^{\pm }$ are listed in Table~\ref{tbl:UpDnA}
of Appendix \ref{sec:PdfTables}.

\subsection{Master equations for calculating uncertainties using the Eigenvector
Basis sets $\{S_{\ell}^{\pm }\}$}
\label{sec:MasterEq}

Let $X(S)$ be any variable that depends on the PDFs. It can be a physical
quantity such as the $W$ production cross section; or a particular PDF such
as the gluon distribution at specific $x$ and $Q$ values; or even one of the
PDF parameters $a_{i}$.  All of these cases will be discussed as examples in
Sections \ref{sec:UncPDF} and \ref{sec:PhysApp}.

The best-fit estimate for $X$ is $X^{0} \! = \! X(S_0)$.
To find the uncertainty, it is only necessary to
evaluate $X$ for each of the $2d$ sets $\{S_{\ell }^{\pm }\}$.
The gradient of $X$ in the $z$-representation can then be calculated,
using a linear approximation that is essential to the Hessian method, by
\begin{equation}
\frac{\partial X}{\partial z_{k}} \, = \,
\frac{X(S_{k}^{+})-X(S_{k}^{-})}{2 \, t}
\label{eq:zGradient}
\end{equation}
where $t$ is the scale used to define $\{S_{\ell }^{\pm }\}$ in
(\ref{eq:eigenvectork-z}). It is useful to define
\begin{eqnarray}
D_{k}(X) &=&X(S_{k}^{+})-X(S_{k}^{-})  \label{eq:Dlength} \\
D(X) &=&\left( \,\sum_{k=1}^d\,[D_{k}(X)]^{2}\right)^{\! 1/2} \\
\widehat{D}_{k}(X) &=&D_{k}(X)/D(X)\;,  \label{eq:Dunit}
\end{eqnarray}
so that $D_{k}(X)$ is a vector in the gradient direction and
$\widehat{D}_{k}(X)$ is the unit vector in that direction.

The gradient direction is the direction in which $X$ varies most rapidly,
so the largest variations in $X$ permitted by (\ref{eq:sphere}) are obtained
by displacement from the global minimum $S_{0}$ in the gradient direction
by a distance $\pm T$.  Hence
\begin{equation}
\Delta X \, = \, \sum_{k=1}^{d} \, (T\,\widehat{D}_{k})
\frac{\partial X}{\partial z_{k}}  \; .
\label{eq:MasterEqDerivation}
\end{equation}
From this, using (\ref{eq:zGradient})--(\ref{eq:Dunit}), we obtain the
\emph{master equation} for calculating uncertainties,
\begin{equation}
\fbox{\mbox{\mbox{$\displaystyle \;
\Delta X  \, = \, \frac{T \strut}{2t \strut} \, D(X)
\; $}}} \; .
\label{eq:MasterEq}
\end{equation}
This equation is applied to obtain numerical results in
Sections \ref{sec:UncPDF} and \ref{sec:PhysApp}.

For applications, it is often important also
to construct the PDF sets $S_{X}^{+}$ and $S_{X}^{-}$
that achieve the extreme values $X = X^{0} \pm \Delta X$.
Their $z$-coordinates are
\begin{equation}
z_{k}(S_{X}^{\pm }) \, = \, \pm \,T\,\widehat{D}_{k}(X) \; ,
\label{eq:zsubipm}
\end{equation}
which follows from the derivation of (\ref{eq:MasterEq}).
Their physical parameters $\{a_i\}$ then
follow from Eqs.(\ref{eq:MikTransform}) and (\ref{eq:DiffEq}):
\begin{equation}
\fbox{\mbox{\mbox{$\displaystyle \;
a_{i}(S_{X}^{\pm })-a_{i}^{0} \, = \,
{\frac{\pm T^{\strut}}{2t_{\strut}}} \,
\sum_{k=1}^d \, \widehat{D}_{k}(X) \,
[a_{i}(S_{k}^{+})-a_{i}(S_{k}^{-})]
\; $}}} \; .
\label{eq:AiSxpm}
\end{equation}
In practice, we calculate the parameters for $S_{X}^{+}$ and $S_{X}^{-}$
by applying (\ref{eq:AiSxpm}) directly to the parton shape parameters
$\, \ln A_0^{u_v}$, $A_1^{u_v}$,\ldots listed in Table~\ref{tbl:UpDnA},
except that the normalization factors $A_{0}^{u_{v}}$, $A_{0}^{d_{v}}$,
and $A_{0}^{g}$ are computed from the momentum sum rule and quark number
sum rules
\begin{equation}
\sum_{i} \int_0^1 \! x \, f_i(x) \, dx \,= \,1
\label{eq:MomentumSumRule}
\end{equation}
\begin{equation}
\int_0^1 \! u_v(x) \, dx \,= \, 2 \; , \qquad
\int_0^1 \! d_v(x) \, dx \,= \, 1
\label{eq:NumberSumRules}
\end{equation}
to ensure that those sum rules are satisfied exactly.




\section{Uncertainties of Parton Distributions}
\label{sec:UncPDF}

\subsection{Uncertainties of the PDF parameters $\{a_{i}\}$}
\label{sec:UncParamAi}

As a useful as well as illustrative application of the general formalism,
let us find the uncertainties on the physical PDF parameters $a_{i}$.
We only need to follow the steps of Sec.~\ref{sec:MasterEq}.
Letting $X=a_{i}$ for a particular $i$,
Eqs.~(\ref{eq:Dlength}) and (\ref{eq:DiffEq}) give
\begin{equation}
D_{k}(a_{i})=\,a_{i}(S_{k}^{+})-a_{i}(S_{k}^{-})=2t\,M_{ik} \; .
\label{eq:zsubi}
\end{equation}
The uncertainty on $a_{i}$ in the global analysis follows from the
master equation (\ref{eq:MasterEq}):
\begin{equation}
\Delta a_{i} \, = \,
T\left( \sum_{k} M_{ik}^{\,2} \right) ^{\, 1/2} \; .
\label{eq:DelAi}
\end{equation}

The parameter sets $\{a_j(a_i^{+})\}$ and $\{a_j(a_i^{-})\}$ that produce
the extreme values of $a_i$ can be found using (\ref{eq:AiSxpm}).
In the conceptual Fig.~\ref{fig:Illustrate}, the parton distribution set
with the largest value of $a_{i}$ for
$\Delta \chi_{\mathrm{global}}^{2}=T^{2}$ is depicted as point
$\mathbf{p(i)}$.

The uncertainties $\{\Delta a_{i}\}$ of the standard parameter set, calculated
from (\ref{eq:DelAi}) with $T=5$ are listed along with the central values
$\{a_{i}^{0}\}$ in Table~\ref{tbl:AiMatrix}.
To test the quadratic approximation, asymmetric errors are also listed.
These are defined by displacements in the gradient direction (\ref{eq:zsubi})
that are adjusted to make $\Delta \chi ^{2}$ exactly equal to
$T^{2}=25$. They agree quite well with the errors calculated using
(\ref{eq:DelAi}), which shows that the quadratic approximation is adequate
for our purposes.

Table~\ref{tbl:AiMatrix} also lists the components of the displacement
vectors $z_{k}(a_{i}^{+})$ of (\ref{eq:zsubipm}), which have been renormalized
to $\sum z_{k}^{\,2}=1$.
These reveal which features of the PDFs are governed most strongly by specific
eigenvector directions. The table is divided into sections according to the
various flavor combinations that are parametrized.  One can see for example
that the flattest direction $z_{16}$ is strongly related to the gluon
parameters $A_{1}^{g}$ and $A_{2}^{g}$, confirming that the gluon
distribution at $Q_{0}=1 \, \mathrm{GeV}$ is a highly uncertain
aspect of the PDFs.  The second-flattest
direction $z_{15}$ relates mainly to the $\bar{d}/\bar{u}$ ratio, as seen
by the large components along $z_{15}$ for $B_{0}^{\bar{d}/\bar{u}}$ and
$B_{1}^{\bar{d}/\bar{u}}$.  Meanwhile, the steepest direction $z_{1}$
mainly influences the valence quark distribution, via $A_{1}^{u_v}$.

All of the parametrized aspects of the PDFs at $Q_0$, namely $u_v$, $d_v$,
$g$, $\bar{d} + \bar{u}$, and $\bar{d} / \bar{u}$ receive substantial
contributions from the four flattest directions $13$--$16$, which shows
that the current global data set could not support the extraction of much
finer detail in the PDFs.  This can be confirmed by noting that the
error ranges of the individual parameters $a_i$ are not small.

\subsection{Uncertainties of the PDFs}
\label{sec:PDFUncertainty}

\figUncGLUONUPQUARK  
\figErrGLUONUPQUARK  
The uncertainty range of the PDFs themselves can also be explored using the
eigenvector method.  For example, letting the gluon distribution $g(x,Q)$ at
some specific values of $x$ and $Q$ be the variable $X$ that is extremized by
the method of Sec.~\ref{sec:MasterEq} leads to the extreme gluon distributions
shown in the left-hand side of Fig.~\ref{fig:UncGLUONUPQUARK}.
The envelope of such
curves, obtained by extremizing at a variety of $x$ values at fixed $Q$,
is shown by the shaded region, which is defined by $T \! = \! 10$,
{\textit i.e.}, by allowing $\chi_{\rm{global}}^2$ up to $100$ above its
minimum value.


The right-hand side of Fig.~\ref{fig:UncGLUONUPQUARK} similarly shows the
allowed region and two specific cases for the $u$-quark distribution.
The uncertainty is much smaller than for the gluon, reflecting the large
amount of experimental data included in the global analysis that is
sensitive to the $u$ quark through the square of its electric charge.

The dependence on $x$ in these figures is plotted as a function of
$x^{1/3}$ to better display the region of current experimental interest.
The values are weighted by a factor $x^{5/3}$, which makes the
area under each curve proportional to its contribution to the momentum
sum rule.
Note that the uncertainty decreases markedly with increasing $Q$ as a
result of evolution.
Also note that the gluon distribution is large and fairly well determined
at smaller $x$ values and large $Q$---the region that will be vital for
physics at the LHC.

Figure \ref{fig:ErrGLUONUPQUARK} displays similar
information for $Q = 10 \, {\rm GeV}$, expressed as the fractional
uncertainty as a function of $\, \log x$.  It shows
that the gluon distribution becomes very uncertain at large $x$, \textit{e.g.},
$x \! > \! 0.25\,$.  (At $x \! > \! 0.6$, where the distribution is
extremely small, the lower envelope of fractional uncertainty begins to
rise. This is an artifact of
the parametrization with $A_3^g = 0$; {\it e.g.}, making the parametrization
more flexible by freeing $A_3^g$ with $A_4^g = 6$ leads to a broader allowed
range indicated by the dotted curves.)

The boundaries of the allowed regions for the PDFs are not themselves possible
shapes for the PDFs, since if a particular distribution is extremely high at
one value of $x$, it will be low at other values.  This can be seen most
clearly in the gluon distributions of Figs.~\ref{fig:UncGLUONUPQUARK} and
\ref{fig:ErrGLUONUPQUARK}, where the extreme PDFs shown push the envelope on
the high side in one region of $x$, and on the low side in another.




\section{Uncertainties of Physical Predictions}
\label{sec:PhysApp}

In applying the Hessian method to study the uncertainties of physical
observables due to PDFs, it is important to understand how the predictions
depend on the tolerance parameter $T$, and how well $T$ can be determined. We
discuss these issues first, and then proceed to illustrate the utility of this
method by examining the predictions for the rapidity distribution of $W$ and $Z$
boson production as well as the correlation of $W$ and $Z$ cross-sections in
$p \bar{p}$ collisions.

First, note that the uncertainties of all predictions are linearly dependent on
the tolerance parameter $T$ in the Hessian approach, by the master formula
(\ref{eq:MasterEq}); hence they are easily scalable. The appropriate value of
$T$ is determined, in principle, by the region of ``acceptable fits'' or
``reasonable agreement with the global data sets'' in the PDF parameter space.
Physical quantities calculated from PDF sets within this region will range over
the values that can be considered ``likely'' predictions.
The language used here is intentionally imprecise, because as discussed in the
introductory sections, the experimental and theoretical input
to the function $\chi_{\mathrm{global}}^{2}$ in the global analysis---in
particular the unknown systematic errors reflected in apparent abnormalities of
some reported experimental errors as well as incompatibilities between
some data sets---makes it difficult to assign an unambiguous value to $T$.
However, as mentioned in Sec.~\ref{sec:Neighborhood}, self-consistency
considerations inherent in our basic assumption that the 15 data sets used in
the global analysis are acceptable and compatible, in conjunction with the
detailed comparison to experiment conducted in LMM \cite{LMM} using the same
$\chi_{\mathrm{global}}^{2}$ function, yield a best estimate of $T \approx 10
\; {\rm to} \; 15$ (Eq.~\ref{eq:Tapprox100}). Details of these considerations
are discussed in Appendix \ref{sec:Tolerance}.

Of the estimates of $T$ described there, the most quantitative one is
based on the algorithm of LMM \cite{LMM} to combine $90$\% confidence level
error bands from the 15 individual data sets for any specific physical variable
such as the total production cross section of $W$ or $Z$ at the Tevatron or LHC.
(\textit{Cf.}\ Appendix A for a summary, and LMM \cite{LMM} for the detailed
analysis.) For the case of $\sigma_W^\mathrm{TeV}$, the uncertainty according
to the specific algorithm is $\pm 4$\%, corresponding to $T \sim 13$. With our
working hypothesis $T \approx 10$ -- $15$, the range of the uncertainty of
$\sigma_W^\mathrm{TeV}$ will be $\pm 3.3\%$ to $\pm 4.9\%$.

The numerical results on applications presented in the following subsections
are obtained with the same choice of $T$ as in Sec.~\ref{sec:UncPDF},
\textit{i.e.}~$T=10$. Bearing in mind the linear dependence of the uncertainties
on $T$, one can easily scale these up by the appropriate factor if a more
conservative estimate for the uncertainty of any of the physical quantities is
desired. We should also note that the experimental data sets used in this
analysis are continuously evolving. Some data sets (cf.\ Table
\ref{tbl:DatSet} in Appendix A) will soon be updated
(H1, ZEUS) or replaced (CCFR).%
\footnote{%
{\it Cf.}\ Talks presented by these collaborations at DIS2000 \emph{Workshop
on Deep Inelastic Scattering and Related Topics}, Liverpool, England, April
2000.\label{fn:DIS2000}} %
In addition, theoretical uncertainties have yet to be systematically
studied and incorporated.  Therefore, the specific results presented in this
paper should be considered more as a demonstration of the method rather than
definitive predictions.  The latter will be refined as new and better inputs
become available.

\subsection{Rapidity distribution for W production}
\label{sec:Wrapidity}

Figure~\ref{fig:DsigDyOneAndTwo} shows the predicted rapidity distribution
$d\sigma/dy$ for $W^+$ production in $p \bar{p}$ collisions at
$\sqrt{s} = 1.8 \, {\rm TeV}$.  The cross section is not symmetric in $y$
because of the strong contribution from the valence $u$-quark in the
proton---indeed, the forward/backward asymmetry produces
an observable asymmetry in the distribution of leptons from $W$ decay, which
provides an important handle on flavor ratios in the current global analysis.

\figDsigDyOneAndTwo  

The left-hand side of Fig.~\ref{fig:DsigDyOneAndTwo} shows the six rapidity
distributions that give the extremes (up or down) of the integrated cross
section $\sigma = \int \frac{d\sigma}{dy} \, dy$, the first moment
$\langle y \rangle$, or the second moment $\langle y^2 \rangle$,
as calculated using the Hessian formalism for $T \! = \! 10$.
To show the differences more clearly, the right-hand side shows
the difference between each of these rapidity distributions and
the Best Fit distribution.

Figure~\ref{fig:DsigDyThree} shows three of the same difference curves
as in Fig.~\ref{fig:DsigDyOneAndTwo} along
with results obtained using the Lagrange Multiplier method of
LMM \cite{LMM}.  The good agreement shows that the Hessian
formalism, with its quadratic approximation (\ref{eq:taylor}),
works well at least for this application.

Figure~\ref{fig:DsigDyFour} shows the same three curves from
Fig.~\ref{fig:DsigDyOneAndTwo}, together
with 6 random choices of the PDFs with
$\Delta \chi_{\mathrm{global}}^2 = 100$.  These random sets were
obtained by choosing random directions in $\{z_i\}$ space and
displacing the parameters from the minimum in those directions until
$\chi_{\mathrm{global}}^2$ has increased by $100$.
Note that none of this small number of random sets give good
approximations to the three extreme curves.  This is not really
surprising, since the extrema are produced by displacements
in specific (gradient) directions; and in 16-dimensional space,
the component of a random unit vector along any specific direction
is likely to be small.  But it indicates that producing large numbers
of random sets would at best be an inefficient way to unearth the
extreme behaviors.
\figDsigDyThree  
\figDsigDyFour  
\clearpage

\subsection{Correlation between W and Z cross sections}
\label{WZCorrelation}

One can ask what are the error limits on two quantities $X$ and
$Y$ simultaneously, according to the
$\Delta \chi_{\mathrm{global}}^2 \! < T^2$ criterion.  In the Hessian
approximation, the boundary of the allowed region is an 
ellipse \cite{Paper0}.  The ellipse can be expressed elegantly in a
``Lissajous figure'' form
\begin{eqnarray}
X &=& X^0 \, + \, \Delta X \, \sin(\theta + \phi) \nonumber
\\
Y &=& Y^0 \, + \, \Delta Y \, \sin(\theta) \; ,
\label{eq:ellipse}
\end{eqnarray}
where $0 < \theta < 2 \pi$ traces out the boundary.  The shape of the
ellipse is governed by the phase angle $\phi$, which is given by the
dot product between the gradient vectors for $X$ and $Y$ in $\{z_k\}$ space:
\begin{equation}
\cos \phi =
\sum_{k=1}^d \, \widehat{D}_k(X) \, \widehat{D}_k(Y) \; ,
\end{equation}
where $\widehat{D}_i(X)$ and $\widehat{D}_i(Y)$ are defined by
(\ref{eq:Dunit}).

As an example of this, $T \! = \! 10$ error
limits for $W^\pm$ and $Z^0$ production at the Tevatron are shown
in Fig.~\ref{fig:Ellipse}.  The error limits on the separate predictions
for these cross sections are each about $3.3\%$ for $T \! = \! 10$.
The predictions
are strongly correlated ($\cos \phi = 0.60$), in part because the same
quark distributions---in different combinations---are responsible
for both $W$ and $Z$ production, and in part because the
uncertainties of all the quark distributions are negatively correlated
with the more uncertain gluon distribution, and hence positively correlated
with each other.

\figEllipse  

The $W$ and $Z$ cross sections from CDF (dashed) and D\O{} (dotted) are also
shown in Fig.~\ref{fig:Ellipse} \cite{CDFD0}.  (The measured quantities
$\sigma_W \cdot B_{W \to e \nu}$ and
$\sigma_Z \cdot B_{Z \to e^{+} e^{-}}$
were converted to $\sigma_W$ and $\sigma_Z$ using world average values
for the branching ratios \cite{PDG}; the measured CDF and D\O{} branching
ratios for $W$ agree with the world average to within about $1\%$.)
The data points are shown in the form of error bars defined by combining
statistical and systematic errors (including the errors in decay branching
ratios) in quadrature.
The errors in these measurements are also highly correlated, in part
through the uncertainty in overall luminosity which both cross
sections are proportional to---so the experimental points would
also be better represented by ellipses.  The two experiments in fact
use different assumptions for the inelastic $p \bar p$ cross section which
measures the luminosity:  CDF uses its own measurement, while D\O{}
uses the world average.
The dot-dashed data point shows the result of reinterpreting the CDF point
by scaling the luminosity down by a factor $1.062$ to correspond to the
world average $p \bar p$ cross section \cite{CDFD0}.




\section{Summary and Concluding Remarks}

\label{sec:Conclude}

Experience over the past two decades has shown that minimizing a suitably
defined $\chi_{\mathrm{global}}^{2}$ is an effective way to extract parton
distribution functions consistent with experimental constraints in the PQCD
framework. The goal of this paper has been to expand the
analysis to make quantitative estimates of the uncertainties of PDFs and their
predictions, by examining the behavior of $\chi_{\mathrm{global}}^{2}$ in the
neighborhood of the minimum.  The techniques developed in Ref.~\cite{Paper0}
allow us to apply the traditional error matrix approach reliably in the
global analysis environment. The eigenvectors of the Hessian (inverse
of the error matrix) play a crucial role, both in the adaptive procedure to
accurately calculate the Hessian itself, and in the derivation of the
master formula (\ref{eq:MasterEq}) for determining the uncertainties of parton
distributions and their predictions.

Our principal results are: (i) the formalism developed in
Sec.~\ref{sec:MasterEq}, leading to the master formulas; and (ii) the Best Fit
parton distribution set $S_0$ plus the $2 d$ Eigenvector Basis sets
$S_{k}^{\pm}$ presented in Sec.~\ref{sec:Eigenvectors}, which are used in
applications of the master formula.  The uncertainties are
proportional to $T$, the tolerance parameter for
$\Delta\chi_{\mathrm{global}}^{2}$. We present several estimates, based on
current experimental and theoretical input, that suggest $T$ is in the range
$10$ -- $15$.  It is important to note, however, that this estimate can, and
should, be refined in the near future. First, several important data sets used
in the global analysis will soon be updated or replaced.
Secondly, there are other
sources of uncertainties which have yet to be studied and included in the
analysis in a full evaluation of uncertainties. (The work of Botje
\cite{Botje} describes possible ways to incorporate some of these.)

This paper, focusing on the presentation of a new formalism and its utility,
represents the first step in a long-term project to investigate the
uncertainties of predictions dependent upon parton distributions. We plan to
perform a series of studies on processes in precision SM measurements (such as
the $W$ mass) and in new physics searches (such as the Higgs production
cross section), which are sensitive to the parton distributions.

\paragraph*{Acknowledgements}

This work was supported in part by NSF grant PHY--9802564. We thank M.\ Botje
and F.\ Zomer for several interesting discussions and for valuable comments on
our work.  We also thank John Collins, Dave Soper, and other CTEQ colleagues
for discussions.

\clearpage

\appendix




\section{Estimates of the Tolerance Parameter for
$\Delta \chi^2_\mathrm{global}$}
\label{sec:Tolerance}

This appendix provides details of the various approaches mentioned in
Sec.~\ref{sec:Neighborhood} and Sec.~\ref{sec:PhysApp} to estimate the
tolerance parameter $T$ defined in Eq.~(\ref{eq:tolerance}). In our global
analysis based on $\Delta \chi_{\mathrm{global}}^{2}$, all uncertainties of
predictions of the PDFs according to the master formula Eq.~(\ref{eq:MasterEq})
are directly proportional to the value of $T$.

The first two estimates rest on considerations of self-consistency which are
required by our basic assumption that the 15 data sets used in the global
analysis (see Table \ref{tbl:DatSet}) are \emph{acceptable} and mutually
\emph{compatible}---in spite of the departure from ideal statistical
expectations exhibited within many of the individual data sets, as well as
apparent incompatibility between experiments when the errors are interpreted
according to strict statistical rules \cite{GKK}.  A third estimate follows
from the analyses in our companion paper LMM \cite{LMM}.
Based on these three estimates, we adopt the range
$T \approx 10 \; {\rm to} \; 15 $ as our working hypothesis,
as was quoted in Eq.~(\ref{eq:Tapprox100}), and
used in Secs.~\ref{sec:UncPDF} and \ref{sec:PhysApp} to obtain the
numerical results shown in the plots.

Finally, it is of interest to compare these estimates of the tolerance
parameter with the traditional---although by now generally recognized
as questionable---gauge provided by differences between published PDFs.

\textbf{1. Tolerance required by acceptability of the experiments}: One can
examine how well the best fit $S_0$ agrees with the individual data sets, by
comparing $\chi_n^2$ in Eq.~(\ref{eq:Chi2global}) with the range $N_n \pm
\sqrt{2N_n}$ that would be the expected $1\,\sigma$ range if the errors were
ideal.  The largest deviations are found to lie well outside that range:
$\chi^2_n \! - \! N_n \, (\sqrt{2N_n}) = 65.5 \, (17.7)$,
$-64.8 \, (18.5)$, $65.1 \, (19.3)$, $-25.9 \, (15.4)$,
$22.4 \, (8.1)$, for experiments $n=2,3,4,10,15$
respectively. By attributing the ``abnormal'' $\chi^2_n$'s to unknown
systematic errors or unusual fluctuations (or both), and accepting them in the
definition of $\chi_{\mathrm{global}}^{2}$ for the global analysis, we must
anticipate a tolerance for the latter which is larger than that for an
``ideal'' $\chi^2$-function. (\textit{Cf.}\ Appendix A of LMM \cite{LMM} for a
quantitative discussion of the increase in $T$ due to neglected systematic
errors.) Since the sources of the deviation of these real experimental errors
from ideal expectations are not known, it is not possible to derive specific
values for the overall tolerance. However, the sizes of the above quoted
deviations (which, in each case, imply a very improbable fit to \emph{any}
theory model, according to ideal statistics) suggest that the required
tolerance value for the overall $\Delta\chi_{\mathrm{global}}^{2}$ (involving
1300 data points) must be rather large.

\tblDatSet

\textbf{2. Tolerance required by mutual compatibility of the experiments}:
We can quantify the degree of compatibility among the 15 data sets by removing
each one of them in turn from the analysis, and observing how much the total
$\chi^{2}$ for the remaining 14 sets can be lowered by readjusting the
$\{a_i\}$.  This is equivalent to minimizing $\chi^{2}$ for each possible
14-experiment subset of the data, and then asking how much increase in the
$\chi^{2}$ for those 14 experiments is necessary to accommodate the return of
the removed set.  These increases are listed as $\Delta_n$ in Table
\ref{tbl:DatSet}. They range up to $\approx \! 20$.  In other words, we have
implicitly assumed that when a new experiment requires an increase of $20$ in
the $\chi_{\mathrm{global}}^{2}$ of a plausible global data set, that new
experiment is nevertheless sufficiently consistent with the global set that it
can be included as an equal partner.\footnote{Since 5 or 6 of the experiments
require $\Delta_n$ in the range of $10$ to $20$, this level of inconsistency
is not caused by problems with just one particular experiment---which would
simply invite the permanent removal of that experiment from the analysis.}
Hence the
value of $T^2$ must be substantially larger than $20$.

\textbf{3. Tolerance calculated from confidence levels of individual
experiments}: In \cite{LMM}, we examine how the quality of fit to each of the
15 individual experiments varies as a function of the predicted value for
various specific observable quantities such as $\sigma_W$ or $\sigma_Z$. The
fit parameters $\{a_i\}$ are continuously adjusted by the Lagrange Multiplier
method to yield the minimum possible value of $\chi_{\mathrm{global}}^{2}$ for
given values of the chosen observable. The constrained fits obtained this
way, interpreted as ``alternative hypotheses'' in statistical analysis,
are then compared to each of the 15 data sets to obtain a $90\%$ confidence
level error range for the individual experiments. Finally, these errors are
combined with a definite algorithm to provide a quantifiable uncertainty
measure for the cross section. In the case of the $W$ production cross section
at the Tevatron, $\sigma_W^\mathrm{TeV}$, this procedure yields an uncertainty
of $\pm 4\%$, which translates into a value of $\approx 180$ for $\Delta
\chi_{\mathrm{global}}^{2}$, or $T \approx 13$. This method is definite, but
it is, in principle, process-dependent. However, when the same analysis is
applied to $\sigma_Z^\mathrm{TeV}$, $\sigma_W^\mathrm{LHC}$ and
$\sigma_Z^\mathrm{LHC}$ (which probe different directions in the PDF parameter
space), we find $\Delta \chi_{\mathrm{global}}^{2}$ to be consistently in the
same range as for $\sigma_W^\mathrm{TeV}$, even though the percentage errors
on the cross section vary from $4$\% at the Tevatron to $10$\% at LHC.

\textbf{4. Comparison of tolerance figures to differences between published
PDFs}: Table \ref{tbl:ChiSq} lists the value obtained when our
$\chi_{\mathrm{global}}^{2}$ is computed using various current and historical
PDF sets.  The $\Delta\chi^2$ column lists the increase over the CTEQ5M1 set.
Typical values for the modern sets are similar to the range $100$ -- $225$
that corresponds to $T \! = \! 10$ -- $15$.  For previous generations of PDF
sets, $\chi_{\mathrm{global}}^{2}$ is much larger---not surprisingly, because
the obsolete sets were extracted from much less accurate data, and without
some of the physical processes such as $W$ decay lepton asymmetry and
inclusive jet production.

\tblChiSq



\section{Tests of the quadratic approximation}
\label{sec:QuadApprox}

The Hessian method relies on a quadratic approximation (\ref{eq:taylor}) to
the effective $\chi^2$ function in the relevant neighborhood of its minimum.
To test this approximation, Fig.~\ref{fig:DchisqrVSzi} shows the dependence
of $\chi^2$ along a representative sample of the eigenvector directions.
The steep directions $1$ and $4$ are indistinguishable from the ideal quadratic
curve $\Delta \chi^2 = z^2$.  The shallower directions $7$, $10$, $13$, $16$,
are represented fairly well by that parabola, although they exhibit
noticeable cubic and higher-order effects. The agreement at small $z$ is not
perfect because we adjust the scale factors $s_k$ in (\ref{eq:vij})
(see footnote \ref{fn:scale}) to improve the average agreement over the
important region $z \lesssim 5$, rather than defining the matrix $H_{ij}$ in
(\ref{eq:taylor}) strictly by the second derivatives at $z=0$.
For this reason, the scale factors $s_k$ in (\ref{eq:eigenvectork-a})
are somewhat different from the $\sqrt{1 / \epsilon_{k}}$ suggested by the
Taylor series:  the flattest directions are extremely flat only over very
small intervals in $z$, so it would be misleading to represent them solely
by their curvature at $z=0$.%
\figDchisqrVSzi  

Figure \ref{fig:DchisqrRandir} %
\figDchisqrRandir  
shows the dependence of $\chi^2$ along some
random directions in $\{z_i\}$ space.  The behavior is reasonably close to
the ideal quadratic curve $\Delta \chi^2 = z^2$, implying that the quadratic
approximation (\ref{eq:taylor}) is adequate.  In particular, the approximation
gives the range of $z$ permitted by $T=10$ to an accuracy of $\approx \!
30\%$. Since the tolerance parameter $T$ used to make the uncertainty estimates
is known only to perhaps $50\%$, this level of accuracy is sufficient.
\clearpage



\section{Table of Best Fit $S_{0}$}
\label{sec:S0Table}

Table~\ref{tbl:AiMatrix} lists the parameter values that define the
``best fit'' PDF set $S_{0}$ which minimizes
$\chi_{\mathrm{global}}^{2}$.  It also lists the uncertainties (for $T=5$)
in those parameters.

For each of the $d \! = \! 16$ parameters, Table~\ref{tbl:AiMatrix} also
lists the components of a unit vector $z_1,\ldots,z_d$ in the eigenvector
basis.  That unit vector gives the direction for which the parameter varies
most
rapidly with $\chi_{\mathrm{global}}^{2}$, {\it i.e.}, the direction along
which the parameter reaches its extreme values for a given increase in
$\chi_{\mathrm{global}}^{2}$.  For parameter $a_i$, the components $z_k$
are proportional to $M_{ik}$ according to Eq.~(\ref{eq:zsubi}).

\tblAiMatrix



\section{Table and Graphs of the Eigenvector sets $S_{\ell}^{\pm}$}
\label{sec:PdfTables}

\tblUpDnA
\tblUpDnB
\figBasisPictures  
Table~\ref{tbl:UpDnA} and its continuation Table~\ref{tbl:UpDnB} completely
specify the PDF Eigenvector Basis sets $S^+_\ell$ and $S^-_\ell$ by listing
all of their parameters at $Q_0$.  The notation and the best-fit setÿ$\, S_0$
are specified at the beginning of the table.

The coefficients listed provide all of the information that is needed for
applications.  For completeness, however, we state here explicitly the
connections between these coefficients and the constructs that were
used elsewhere in the paper to derive them.
The fit parameters $\{a_i\}$ are related to the tabulated parameters by
\begin{equation}
\begin{array}{llll}
a_{1}=A_{1}^{u_{v}}+1, &
a_{2}=A_{2}^{u_{v}}, &
a_{3}=\ln (1+A_{3}^{u_{v}}), &
a_{4}=A_{4}^{u_{v}}, \\
a_{5}=A_{2}^{d_{v}}, &
a_{6}=\ln (1+A_{3}^{d_{v}}), &
a_{7}=A_{4}^{d_{v}}, &
a_{8}=\ln {\widetilde{f_{g}}}, \\
a_{9}=A_{1}^{g}+1, &
a_{10}=A_{2}^{g}, &
a_{11}=A_{1}^{\,\bar{d}+\bar{u}}+1,&
a_{12}=A_{2}^{\,\bar{d}+\bar{u}}, \\
a_{13}=\ln (1+A_{3}^{\bar{d}+\bar{u}}), &
a_{14}=\ln B_{0}^{\bar{d}/\bar{u}}, &
a_{15}=B_{1}^{\,\bar{d}/\bar{u}}, &
a_{16}=B_{4}^{\,\bar{d}/\bar{u}}
\end{array}
\;.  \label{eq:alist}
\end{equation}

Each of the $a_i$ is thus related to a single PDF parameter, except
for $a_8$ which is related to $\widetilde{f_g}$, the momentum fraction
carried by gluons, and is thus determined by $A_0^{g},\ldots,A_4^{g}$.
The matrix elements of the transformation from the $a_i$ coordinates
to the eigenvector coordinates are given by
\begin{equation}
M_{i\ell} = \frac{a_i(S^{+}_\ell) - a_i(S^{-}_\ell)}{2 \, t}
\end{equation}
according to (\ref{eq:DiffEq}), where $t = 5$ because that value
was used to generate the $S_{\ell}^{\pm}$.
Eqs.~(\ref{eq:orthonormal}) and (\ref{eq:Mik}) imply
\begin{equation}
\sum_{i=1}^d \, M_{i\ell}\,M_{ik} = s_\ell \, s_k \, \delta_{\ell k} \;.
\end{equation}
For $\ell \ne k$, this becomes $\sum_{i=1}^d M_{i\ell} M_{ik} = 0$, which
can serve as a check on numerical accuracy; while for $\ell = k$,
it becomes $\sum_{i=1}^d M_{i\ell}^{\, 2} = s_{\ell}^{\, 2}$ which can be
used to reconstruct $s_1,\ldots,s_d$.

Finally, for the benefit of the reader who is curious about them, graphs are
shown in Fig.~\ref{fig:BasisPictures} of the differences described by each of
the PDF eigenvector sets.  One sees that the steeper directions
(small values of $\ell$) mainly control aspects of the quark distribution,
while the shallower directions (high values of $\ell$) control the gluon
distribution, whose absolute uncertainty is larger.  The variations in the
gluon distribution show less variety than the quarks because the
gluon distribution is described by only $3$ parameters (including
normalization), such that the most general variation for it is
of the form $\frac{\Delta g}{g} = c_1 + c_2\log x + c_3 \log (1-x)$.
\clearpage


\end{document}